\documentclass[%
superscriptaddress,
preprint,
 amsmath,amssymb,
 aps,
]{revtex4-2}

\usepackage{graphicx}
\usepackage{dcolumn}
\usepackage{bm}
\usepackage{multirow}
\usepackage[version=3]{mhchem}
\usepackage{fancyhdr}
\usepackage{amsmath,esint}
\usepackage{verbatim}
\usepackage{spverbatim}
\RequirePackage[normalem]{ulem}
\RequirePackage{color}\definecolor{RED}{rgb}{1,0,0}\definecolor{BLUE}{rgb}{0,0,1}\definecolor{GREEN}{rgb}{0,1,0}

\begin{document}


\title{Length-gauge optical matrix elements in WIEN2k}

\author{Oleg Rubel}%
\email[O.R. email: ]{rubelo@mcmaster.ca, ORCID: 0000-0001-5104-5602}
\affiliation{Department of Materials Science and Engineering, McMaster University, 1280 Main Street West, Hamilton, Ontario L8S 4L8, Canada}

\author{Peter Blaha}
\email[P.B. email: ]{peter.blaha@tuwien.ac.at, ORCID: 0000-0001-5849-5788}
\affiliation{Institute of Materials Chemistry, Vienna University of Technology, Getreidemarkt 9/165-TC, A-1060 Vienna, Austria}

\date{\today}

\begin{abstract}
Hybrid exchange-correlation functionals provide superior electronic structure and optical properties of semiconductors or insulators as compared to semilocal exchange-correlation potentials due to admixing a portion of the non-local exact exchange potential from a Hartree-Fock theory. Since the non-local potential does not commute with the position operator, the momentum matrix elements do not fully capture the oscillator strength, while the length-gauge velocity matrix elements do. So far, length-gauge velocity matrix elements were not accessible in the all-electron full-potential WIEN2k package. We demonstrate the feasibility of computing length-gauge matrix elements in WIEN2k for a hybrid exchange-correlation functional based on a finite difference approach. To illustrate the implementation we determined matrix elements for optical transitions between the conduction and valence bands in GaAs, GaN, (CH$_3$NH$_3$)PbI$_3$ and a monolayer MoS$_2$. The non-locality of the Hartree-Fock exact exchange potential leads to a strong enhancement of the oscillator strength as noticed recently in calculations employing pseudopotentials [Laurien and  Rubel: arXiv:2111.14772 (2021)]. We obtained an analytical expression for the enhancement factor in terms of the difference in eigenvalues not captured by the kinetic energy. It is expected that these results can also be extended to other non-local potentials, e.g., a many-body $GW$ approximation.
\end{abstract}

\maketitle


This paper is dedicated to the $80^{\text{th}}$ birthday of Professor~Karlheinz Schwarz, the founder of the WIEN2k DFT package.

\section{Introduction}\label{sec:Introduction}

Calculations of linear optical properties of solids require matrix elements for electric dipole transitions. Momentum matrix elements
\begin{equation}\label{eq:p_mn velocity gauge}
    \mathbf{p}_{mn}(\mathbf{k}) 
    = \langle m,\mathbf{k} | -i \nabla_{\textbf{r}}  | n,\mathbf{k} \rangle
\end{equation}
are widely used in full-potential codes with periodic boundary conditions~\cite{Ambrosch-Draxl_CPC_175_2006} when optical properties are computed with local potentials (e.g., LDA\footnote{The full list of abbreviations can be found at the end of the paper.} or GGA XC functionals) and referred in the literature as a velocity gauge. (Atomic units will be used throughout the paper.)

\citet{Starace_PRA_3_1971} emphasised the limitations of Eq.~\eqref{eq:p_mn velocity gauge} when representing  matrix elements for electric dipole transitions. Instead, the more general velocity matrix elements should be used
\begin{equation}
    \mathbf{v}_{mn}(\mathbf{k})
    = \langle m,\mathbf{k} | i[\hat{H},\mathbf{r}]  | n,\mathbf{k} \rangle
\end{equation}
with the velocity operator~\cite{Starace_PRA_3_1971} 
\begin{equation}\label{eq:vel operator}
    \hat{\mathbf{v}} = i[\hat{H},\mathbf{r}] = \hat{\mathbf{p}} + i[\hat{V}_{\text{NL}}(\mathbf{r},\mathbf{r}'),\mathbf{r}],
\end{equation}
which contains an additional commutator term $[\hat{V}_{\text{NL}}(\mathbf{r},\mathbf{r}'),\mathbf{r}]$ to account for non-local potentials (e.g., the Hartree-Fock exchange in hybrid XC functionals). With $| n,\mathbf{k} \rangle$ and $E_n(\mathbf{k})$ being eigenstates of the Hamiltonian $\hat{H}$, the alternative (length gauge) matrix elements can be expressed as~\cite{Starace_PRA_3_1971}
\begin{equation}\label{eq:v_mn length gauge general}
    \langle m,\mathbf{k} | i[\hat{H},\mathbf{r}]  | n,\mathbf{k} \rangle
    = i[E_m(\mathbf{k}) - E_n(\mathbf{k})] \langle m,\mathbf{k} | \mathbf{r}  | n,\mathbf{k} \rangle.
\end{equation}
Since the position operator is not well defined for periodic systems, the following substitution is used instead~\cite{Deslippe_CPC_183_2012,Laskowski_PRB_85_2012}
\begin{equation}\label{eq:r approx}
    \mathbf{r} = \lim_{q \rightarrow 0} (e^{iq \mathbf{r}}-1)/iq,
\end{equation}
which leads to a practical expression for the velocity matrix elements in the long wavelength limit~\cite{Rohlfing_PRB_62_2000,Rhim_PRB_71_2005}
\begin{equation}\label{eq:v_mn length gauge FD}
    v_{mn}^{(\alpha)}(\mathbf{k}) = 
    \lim_{q \rightarrow 0} \frac{1}{q}
    \langle m,\mathbf{k}+\mathbf{q}_{\alpha} | \mathrm{e}^{i \mathbf{q}_{\alpha} \cdot \mathbf{r}}  | n,\mathbf{k} \rangle
    [E_{m}(\mathbf{k}+\mathbf{q}_{\alpha})-E_{n}(\mathbf{k})].
\end{equation}
Here $\alpha=x,y,z$ is a Cartesian direction, $m$ and $n$ are band indices, and $\mathbf{q}_{x}=q\,\hat{\mathbf{x}}$, where $\hat{\mathbf{x}}$ is a unit vector in the direction of the $x$ axis.

WIEN2k~\cite{Blaha_WIEN2k_2018,Blaha_JCP_152_2020} is one of the most used full-potential all-electron DFT codes for solids. It offers many XC functionals to open the band gap, including hybrids with a non-local Hartree-Fock potential~\cite{Tran_PRB_83_2011}. However, so far, WIEN2k has implemented only momentum matrix elements to compute optical properties as a part of the \texttt{optic} module~\cite{Ambrosch-Draxl_CPC_175_2006}. \citet{Laurien_arXiv_2021} showed that neglecting the second term in Eq.~\eqref{eq:vel operator} when using hybrid functionals can lead to an underestimation of the squared magnitude of matrix elements for electric dipole transitions between conduction and valence band edges by ca.~30\%.

Here we present a scheme for the calculation of the length-gauge optical matrix elements in WIEN2k based on a finite difference Eq.~\eqref{eq:v_mn length gauge FD} with the help of overlap matrix elements

\begin{equation}\label{eq:M_mn}
    M_{mn}(\mathbf{k},\mathbf{q})
    = \langle u_{\mathbf{k},n} | u_{\mathbf{k}+\mathbf{q},m} \rangle
    \equiv \langle \psi_{\mathbf{k},n} | \mathrm{e}^{-i \mathbf{q} \cdot \mathbf{r}} | \psi_{\mathbf{k}+\mathbf{q},m} \rangle
\end{equation}

\noindent that come from the \texttt{wien2wannier} module~\cite{Kunes_CPC_181_2010}.  This development opens an avenue for the calculation of optical properties (frequency-dependent dielectric tensor, absorption spectrum, optical conductivity, refractive index, reflectively, loss function) in the independent particle approximation with hybrid functionals in WIEN2k.

\section{Methods}

\begin{figure}[t]
\includegraphics{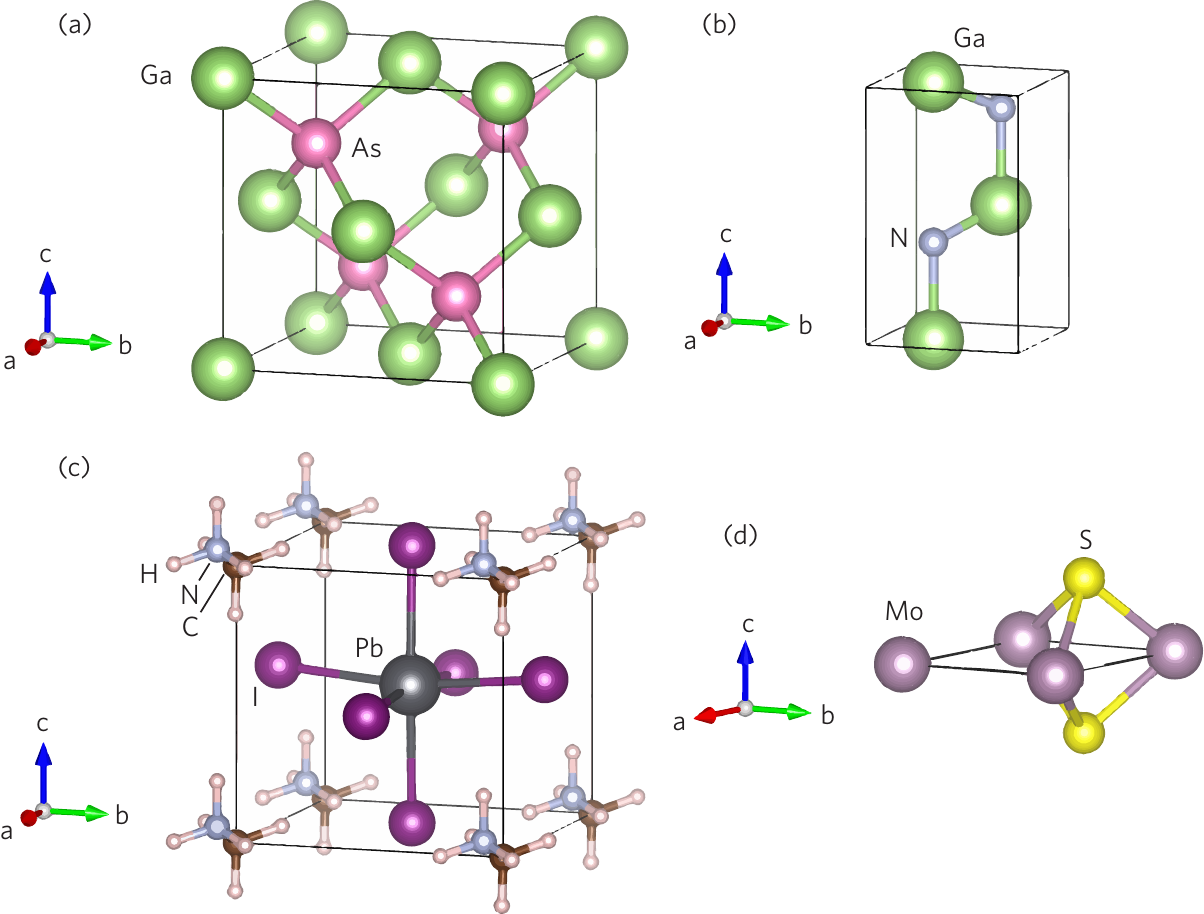}
\caption{Crystal structures: \textbf{(a)} GaAs, \textbf{(b)} GaN, \textbf{(c)} quasi-cubic (CH$_3$NH$_3$)PbI$_3$, and \textbf{(d)} monolayer MoS$_2$.}\label{fig-structures}
\end{figure}

DFT~\cite{Hohenberg_PR_136_1964,Kohn_PR_140_1965} calculations were performed with the WIEN2k package (version 21.1) and the Yukawa screened hybrid (YSH) functional~\cite{Tran_PRB_83_2011}. It was shown that by choosing an appropriate screening length $\lambda$ in the Yukawa potential the YSH functional gives very similar results as the common HSE06 XC functional~\cite{Heyd_JCP_118_2003,Krukau_JCP_125_2006}. Important parameters are summarized in Table~\ref{tab:WIEN2k paramaters}. Experimental structural parameters were used for all solids (Fig.~\ref{fig-structures}) with internal atomic positions optimized at the PBE level when permitted by symmetry. Spin-orbit coupling (SOC) was included in all calculations. The structure of (CH$_3$NH$_3$)PbI$_3$ was represented by a pseudo-cubic cell taken from Ref.~\citenum{Demchenko_PRB_94_2016}, scaled to experimental lattice parameters at 350~K~\cite{Dang_CEC_17_2015,Whitfield_SR_6_2016}, followed by a subsequent relaxation of atomic positions while retaining the experimental lattice parameters. The pseudo-cubic structure means that the following constrains $a=b=c$, $\alpha = \beta = \gamma = 90^{\circ}$ are applied to lattice parameters, while the formal symmetry of the structure (spacegroup P1) is not cubic.

Velocity-gauge optical matrix elements $\mathbf{p}_{mn}(\mathbf{k})$ were calculated using the \texttt{optic} module~\cite{Ambrosch-Draxl_CPC_175_2006} in WIEN2k. Length-gauge optical matrix elements $\mathbf{v}_{mn}(\mathbf{k})$ were obtained with the forward
\begin{equation}\label{eq:v_mn length gauge FD forward}
    v_{mn}^{(\alpha)}(\mathbf{k}) \approx 
    \frac{1}{q}
    \langle u_{\mathbf{k},n} | u_{\mathbf{k}+\mathbf{q}_{\alpha},m} \rangle
    [E_{m}(\mathbf{k}+\mathbf{q}_{\alpha})-E_{n}(\mathbf{k})]
\end{equation}
and central
\begin{equation}\label{eq:v_mn length gauge FD central}
    v_{mn}^{(\alpha)}(\mathbf{k}) \approx 
    \frac{1}{q}
    \langle u_{\mathbf{k}-\frac{1}{2}\mathbf{q}_{\alpha},n} | u_{\mathbf{k}+\frac{1}{2}\mathbf{q}_{\alpha},m} \rangle
    [E_{m}(\mathbf{k}+\frac{1}{2}\mathbf{q}_{\alpha})-E_{n}(\mathbf{k}-\frac{1}{2}\mathbf{q}_{\alpha})]
\end{equation}
finite difference method. The overlap matrix elements $\langle u_{\mathbf{k},n} | u_{\mathbf{k'},m} \rangle$ between the cell-periodic parts of the Bloch functions were generated by the \texttt{wien2wannier} module~\cite{Kunes_CPC_181_2010} (\texttt{case.mmn} output file). The length-gauge optical matrix elements for GaAs computed with YSH  were additionally verified using VASP~\cite{Kresse_PRB_54_1996}, HSE06 and projector augmented-wave potentials~\cite{Kresse_PRB_54_1996,Kresse_PRB_59_1999,Blochl_PRB_50_1994}. Sample scripts that illustrate a detailed workflow can be found in the supporting information section.

The logarithmic percent change
\begin{equation}\label{eq:delta log}
    \Delta =
    \ln \left(
    \frac{\sum|v_{mn}^{(\alpha)}|^2}{\sum|p_{mn}^{(\alpha)}|^2}
    \right) 100\%
\end{equation}
was used to evaluate differences between matrix elements. This approach has the following advantages: (i) does not require a reference, (ii) is more suitable for large changes (greater than a few percents), (iii) it has additive properties, and (iv) in the limit of small changes it reduces to the classical ratio of the relative change to the reference.

\section{Results}

\subsection{Finite difference calibration and validation for local XC}

\begin{figure}[t]
\includegraphics{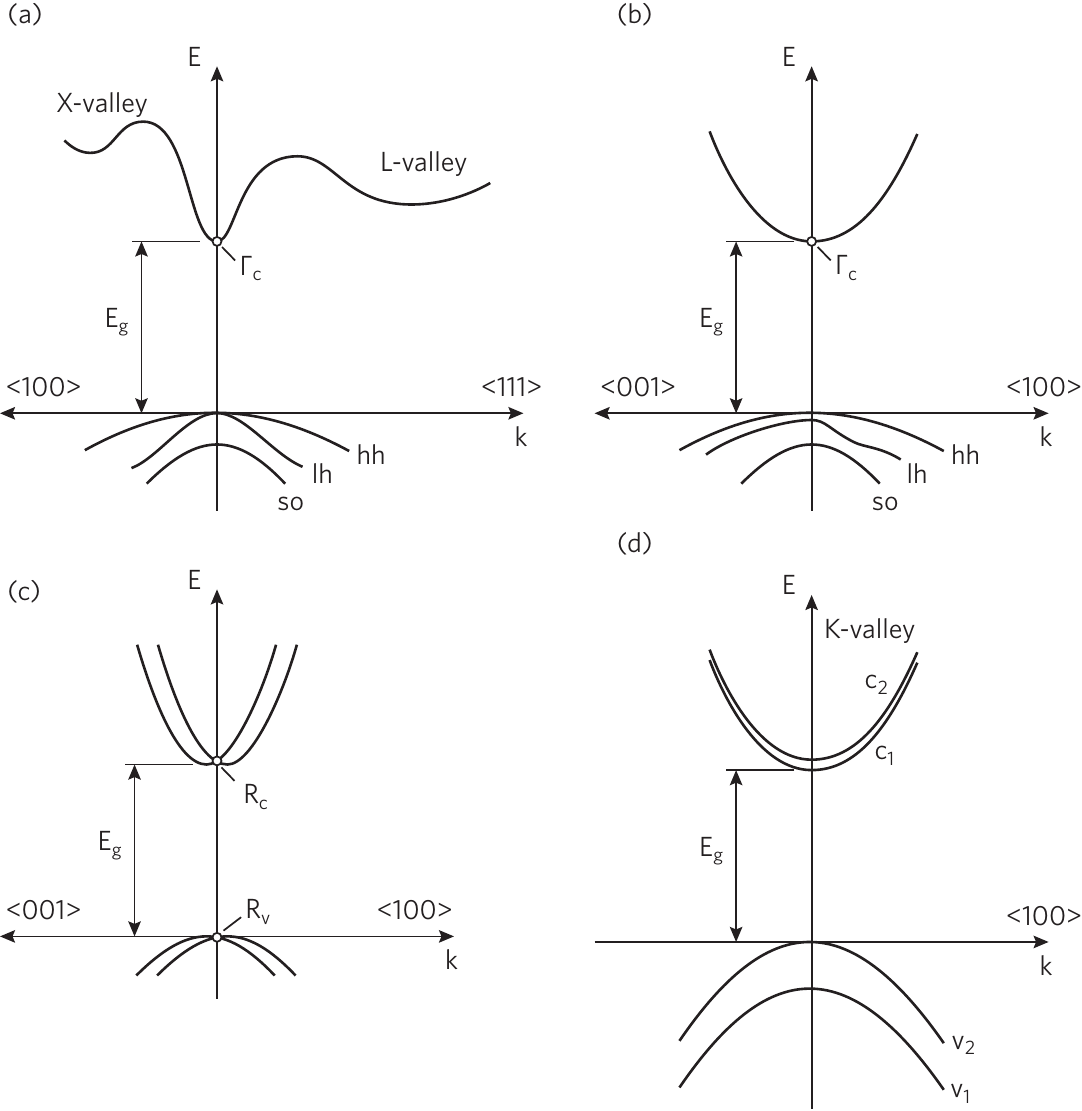}
\caption{Schematic band structures of materials studied with SOC: \textbf{(a)} GaAs, \textbf{(b)} GaN, \textbf{(c)} quasi-cubic (CH$_3$NH$_3$)PbI$_3$, and \textbf{(d)} monolayer MoS$_2$. The band indices 'c', 'hh', 'lh', and 'so' stand for conduction, valence heavy-hole, light-hole, and split-off bands, respectively. The scale of band splittings is exaggerated.}\label{fig-GaAs-schematic-band-struct}
\end{figure}

We selected GaAs and the local PBE XC potential~\cite{Perdew_PRL_77_1996} to prove the feasibility of computing the length-gauge optical matrix elements using the finite difference methods given by Eqs.~\eqref{eq:v_mn length gauge FD forward} and \eqref{eq:v_mn length gauge FD central}. The local potential is selected here since both the length- and velocity-gauge should lead to identical results under these circumstances. It is also important to get a feeling for the step size $q$ at which the finite difference approximation converges to the accurate result given by the momentum matrix element. Here GaAs serves as an important benchmark since the conduction band and light holes are very sharp and non-parabolic (see Fig.~\ref{fig-GaAs-schematic-band-struct}(a)) .

Results presented in Table~\ref{tab:v_mn GaAs PBE} suggest that both, the forward and the central finite differences reproduce the values of the momentum matrix element within 3\% error. The central finite difference converges faster (at the wave vector shift of $q \approx 0.003$~rad~bohr$^{-1}$) and will be used to derive matrix elements for the remaining part of this paper. The numerical noise of the finite difference starts to show up at $q < 10^{-6}$~rad~bohr$^{-1}$.

\subsection{Validation for non-local XC}

After validating our approach with the local potential, we apply it to the non-local YSH XC functional. Again we evaluate the velocity-gauge (momentum) matrix elements and length-gauge (velocity) matrix elements in GaAs. Now we do not expect the two matrix elements to agree given the arguments presented in Sec.~\ref{sec:Introduction}. To cross-check our results, we also computed the velocity matrix elements with VASP, which should be comparable with our $v_{mn}^{(\alpha)}$ values.

Our YSH calculations for GaAs give a band gap of $E_{\text{g}}^{\text{YSH}}=1.24$~eV vs $E_{\text{g}}^{\text{exp}}=1.52$~eV~\cite{smat2017:smi_systemId_1_propertyId_band+gap_database_semiconductor} and previously reported $E_{\text{g}}^{\text{HSE06}}=1.33$~eV~\cite{Kim_PRB_82_2010}, which is a significant improvement over PBE. The band structure is shown schematically in Fig.~\ref{fig-GaAs-schematic-band-struct}(a) where bands are labeled according to the convention. Results presented in Table~\ref{tab:v_mn GaAs HSE} confirm the agreement between WIEN2k and VASP for length-gauge matrix elements within less than 2\% deviation. The total length-gauge oscillator strength between valence and conduction band corresponds to $m_0\sum|v_{cv}^{(x)}|^2 \approx 21$~eV, which agrees well with 20~eV quoted by \citet[sec.~2.6]{Yu_Fundament-semicond} for III-V semiconductors. The momentum (velocity-gauge) matrix elements significantly underestimate the strength of optical transitions, which has been previously reported and quantified in Ref.~\citenum{Laurien_arXiv_2021}. The values of $|p_{cv}|^2$ are almost identical to those obtained with PBE (Table~\ref{tab:v_mn GaAs PBE}), even though the momentum matrix elements were derived from  YSH wave functions.

\subsection{Illustrative applications}

In the previous subsection we showed that calculations of optical properties for GaAs with the non-local hybrid XC functional (YSH or HSE06) requires length-gauge optical matrix elements. If the momentum matrix elements are used instead, the strength of optical transitions is underestimated by 23\%. Next, we show that a similar enhancement of the strength of direct optical transitions is also observed in other semiconductors, such as GaN, (CH$_3$NH$_3$)PbI$_3$, and monolayer MoS$_2$. The corresponding band structures are shown schematically in Fig.~\ref{fig-GaAs-schematic-band-struct}(b$-$d).

The band gap of GaN is well reproduced with YSH: $E_{\text{g}}^{\text{YSH}}=3.19$~eV vs $E_{\text{g}}^{\text{exp}}=3.30$~eV~\cite{smat2017:smi_systemId_15_propertyId_band+gap_database_semiconductor}. As compared to GaAs, optical matrix elements in Table~\ref{tab:v_mn GaN HSE} show an even larger disparity between the length-gauge velocity and the momentum matrix elements.

The monolayer MoS$_2$ has a direct band gap at the $K=(1/3,1/3,1/3)$ point. Due to large excitonic effects~\cite{Hill_NL_15_2015}, direct comparison of the YSH band gap $E_{\text{g}}^{\text{YSH}}=2.22$~eV with experiment is not possible. Thus we use a many-body result $E_{\text{g}}^{G_0 W_0}=2.53$~eV~\cite{Haastrup_2M_5_2018,Gjerding_2M_8_2021} as a reference. Similarly to other materials, the monolayer MoS$_2$ shows a strong enhancement of the matrix elements (Table~\ref{tab:v_mn MoS2 HSE}) with the YSH XC functional. Spin selection rules disable half of the in-plane $v_{cv}^{(x)}$ matrix elements, while the out-of-plane matrix elements $v_{cv}^{(z)}$ are zero for transitions at the band edges due to symmetry arguments.

The pseudo-cubic (CH$_3$NH$_3$)PbI$_3$ has a direct gap at $R=(1/2,1/2,1/2)$. The calculated band gap $E_{\text{g}}^{\text{YSH}}=1.03$~eV is an improvement relative to the PBE band gap (0.46~eV), but it is still far from the experimental $1.5-1.6$~eV for the tetragonal phase~\cite{Ishihara_JL_6061_1994,Dittrich_JPCC_119_2015}. This underestimation is due to the lack of stochastic thermal distortions of the PbI$_6$ octahedra, which further opens the gap by ca. 0.5~eV at room temperature~\cite{Wiktor_JPCL_8_2017,Zheng_PRM_2_2018}. Table~\ref{tab:v_mn MAPbI3 HSE} captures the matrix elements and their anisotropy caused by a reduced (pseudo-cubic) symmetry of the unit cell. Among all materials studied here, this material shows the lowest enhancement of the velocity matrix elements as compared to the momentum matrix elements.

It should be mentioned that momentum matrix elements calculated with the \texttt{optic} module in the presence of SOC have an inaccuracy that progressively increases for heavier elements. The discrepancy between $|p_{mn}^{(x,y,z)}|^2$ values calculated at the PBE level (including SOC) with the \texttt{optic} module and using the finite difference overlap matrix reached ca.~12\% in the case of (CH$_3$NH$_3$)PbI$_3$. The discrepancy fully vanishes when SOC is excluded. After crosschecking the matrix elements with VASP we can conclude that the finite difference results are correct. Since the \texttt{optic} module overestimated $|p_{mn}^{(x,y,z)}|^2$ values at PBE with SOC, the same applies to YSH with SOC results presented in Table~\ref{tab:v_mn MAPbI3 HSE}. After including this error, the true enhancement of YSH matrix elements for (CH$_3$NH$_3$)PbI$_3$ should be about 22\% (10\% average enhancement in Table~\ref{tab:v_mn MAPbI3 HSE} and 12\% \texttt{optic} error for this material). Additional calculations with VASP and HSE06 XC functional with SOC produced a very similar result (23\% enhancement of the matrix elements).

\section{Discussion}

YSH length-gauge $|v_{m,n}^\text{YSH}|^2$ matrix elements are systematically greater than the momentum matrix elements $|p_{m,n}^\text{YSH}|^2$. The enhancement ranges from 22 to 36\% in the following order: (CH$_3$NH$_3$)PbI$_3$, GaAs, GaN, and MoS$_2$ (from the smaller to higher enhancement). This trend prompts a hypothesis that the enhancement is related to the localization of states involved in the optical transition. (CH$_3$NH$_3$)PbI$_3$ has the most extended 5p-I and 6p-Pb states, while MoS$_2$ has the most localized 4d-Mo and 3p-S states at the band edges.

To get further insight on the difference between $|p_{m,n}^\text{YSH}|^2$ and $|v_{m,n}^\text{YSH}|^2$ we write the momentum matrix element in the length gauge. The corresponding operator is expressed as a commutator
\begin{equation}\label{eq:p operator as commutator}
    \hat{\mathbf{p}} = i[\hat{T},\mathbf{r}],
\end{equation}
where $\hat{T}$ is the kinetic energy operator. Following the same logic that leads to Eq.~\eqref{eq:v_mn length gauge general}, we derive an equivalent expression for the momentum matrix element in the length-gauge
\begin{equation}\label{eq:p_mn length gauge general}
    \mathbf{p}_{m,n}(\mathbf{k})
    = i[T_m(\mathbf{k}) - T_n(\mathbf{k})] \langle m,\mathbf{k} | \mathbf{r}  | n,\mathbf{k} \rangle.
\end{equation}
After dividing Eq.~\eqref{eq:v_mn length gauge general} by \eqref{eq:p_mn length gauge general} we obtain
\begin{equation}\label{eq:v_mn renormalization}
    \mathbf{v}_{m,n}(\mathbf{k}) = \mathbf{p}_{m,n}(\mathbf{k})\, \frac{E_m(\mathbf{k}) - E_n(\mathbf{k})}{T_m(\mathbf{k}) - T_n(\mathbf{k})}.
\end{equation}
Thus the 10 to 36\% enhancement of the absolute squared magnitude of velocity matrix elements vs momentum matrix elements in calculations with YSH is directly related to the difference in eigenvalues not captured by the kinetic energy. In contrast, we expect the difference in eigenvalues be fully captured by the kinetic energy, i.e., $[E_m(\mathbf{k}) - E_n(\mathbf{k})]/[T_m(\mathbf{k}) - T_n(\mathbf{k})]=1$, when a local potential is employed. Interestingly, Eq.~\eqref{eq:v_mn renormalization} predicts an \textit{isotropic} renormalization factor shared by all Cartesian directions ($\alpha=x,y,z$). Indeed, materials with anisotropic $v_{mn}^{(\alpha)}$---GaN (Table~\ref{tab:v_mn GaN HSE}) and (CH$_3$NH$_3$)PbI$_3$ (Table~\ref{tab:v_mn MAPbI3 HSE})---show a material-dependent yet isotropic enhancement factor, which is an indirect prove of the validity of Eq.~\eqref{eq:v_mn renormalization}.

The renormalization of momentum matrix elements should have implications for optical properties calculated with non-local potentials (hybrid or quasi-particle $GW$).  The velocity matrix elements enter the frequency-dependent dielectric tensor (an imaginary part of the inter-band contribution) that takes the following form in the independent-particle approximation~\cite{Rohlfing_PRB_62_2000}
\begin{equation}\label{eq:Im[eps] velocity}
    \epsilon''_{\alpha\beta}(\omega) 
    \propto \sum_{v,c} \int_{\mathbf{k}\in \text{BZ}} 
    \frac{v^{(\alpha)}_{v,c}(\mathbf{k})v^{(\beta)}_{c,v}(\mathbf{k})}{\omega^2}
    \, \delta[E_c(\mathbf{k}) - E_v(\mathbf{k}) - \omega]~d\mathbf{k}.
\end{equation}
However, length-gauge $\mathbf{v}_{c,v}$ matrix elements are more difficult to compute than $\mathbf{p}_{c,v}$, especially at the quasi-particle $GW$ level of theory where the finite difference method seems the only available technique~\cite{Deslippe_CPC_183_2012}. Equation~\eqref{eq:v_mn renormalization} opens a convenient possibility to use renormalized momentum matrix elements instead
\begin{equation}\label{eq:Im[eps] momentum}
    \epsilon''_{\alpha\beta}(\omega) 
    \propto \sum_{v,c} \int_{\mathbf{k}\in \text{BZ}}
    \frac{p^{(\alpha)}_{v,c}(\mathbf{k})p^{(\beta)}_{c,v}(\mathbf{k})}{[T_c(\mathbf{k}) - T_v(\mathbf{k})]^2}
    \, \delta[E_c(\mathbf{k}) - E_v(\mathbf{k}) - \omega]~d\mathbf{k} ,
\end{equation}
provided eigenstates are consistent with the potential, and their kinetic energy is known. The last expression should be valid not only for hybrid XC functionals but also for the quasi-particle $GW$ level of theory.

Finally, we would like to comment on the renormalization of optical transition matrix elements proposed by \citet{Levine_PRL_63_1989}
\begin{equation}\label{eq:v_mn renormalization Levine-Allan}
    v_{v,c}^{GW} = v_{v,c}^\text{LDA/PBE} \frac{(E_c - E_v)^{GW}}{(E_c - E_v)^\text{LDA/PBE}}
\end{equation}
that is further used in the literature~\cite{Adolph_PRB_53_1996,Rohlfing_PRB_62_2000}. If we apply Eq.~\eqref{eq:v_mn renormalization Levine-Allan} to the $\Gamma_{so} - \Gamma_{c}$ transition in GaAs, one would expect the absolute squared magnitude of the velocity matrix element to increase by the ratio of $[(E_{c}^\text{YSH}-E_{so}^\text{YSH})/(E_{c}^\text{PBE}-E_{so}^\text{PBE})]^2$ which amounts to +151\%.  This result contradicts the +23\% difference between $|p_{so,c}^\text{PBE}|^2$ and $|v_{so,c}^\text{YSH}|^2$ we observe (compare Tables~\ref{tab:v_mn GaAs PBE} and \ref{tab:v_mn GaAs HSE}). At the same time, the dipole matrix element $|\langle \Gamma_{so} | \mathbf{r}  | \Gamma_{c} \rangle|^2$ becomes 127\% smaller in YSH relative to PBE and counterbalances (partly) the effect of the gap opening. We further identified contributions of the muffin-tin spheres and of the interstitial volume to the value of the dipole matrix element $\langle \Gamma_{so} | \mathbf{r}  | \Gamma_{c} \rangle$ at PBE and YSH levels of theory: 25\% Ga, 44\% As, and 31\% interstitial. All contributions are in phase with each other, and the proportions remain unchanged from PBE to YSH.  Equation~\eqref{eq:v_mn renormalization Levine-Allan}, in contrast, implies the equality of dipole matrix elements $\langle m,\mathbf{k} | \mathbf{r}  | n,\mathbf{k} \rangle^\text{LDA/PBE} = \langle m,\mathbf{k} | \mathbf{r}  | n,\mathbf{k} \rangle ^{GW}$ (see Eq.~\eqref{eq:v_mn length gauge general} and note that $v_{m,n}^\text{LDA/PBE} = p_{m,n}^\text{LDA/PBE} \approx p_{m,n}^{GW}$~\cite{Laurien_arXiv_2021}) leading to a gross overestimation of $v_{m,n}^{GW}$ matrix elements making them inconsistent with the band curvature~\cite{Laurien_arXiv_2021}.

\section{Conclusions}

Strong material-dependent enhancement of the oscillator strength ($22-36$\% in the absolute squared magnitude) is observed in electronic structure calculations of semiconductors with a hybrid XC functional. The origin of the enhancement is traced to the non-local Hartee-Fock exchange potential. The enhancement of the absolute squared magnitude of velocity matrix elements $|v_{m,n}^\text{YSH}|^2$ vs momentum matrix elements $|p_{m,n}^\text{YSH}|^2$ in calculations with non-local potentials is directly related to the difference in eigenvalues not captured by the kinetic energy, i.e., $[E_m(\mathbf{k}) - E_n(\mathbf{k})]^2/[T_m(\mathbf{k}) - T_n(\mathbf{k})]^2$. This enhancement is isotropic and can be readily included in a calculation of the dielectric function. Our enhancement factor is much more accurate than that previously proposed by Levine and Allan $(E_\text{g}^{GW}/E_\text{g}^\text{LDA})^2$, which leads to nonphysically large  $v_{m,n}^{GW}$ matrix elements.

\begin{acknowledgments}
O.R. acknowledges funding provided by the Natural Sciences and Engineering Research Council of Canada under the Discovery Grant Programs RGPIN-2020-04788. Calculations were performed using the Compute Canada infrastructure supported by the Canada Foundation for Innovation under John R. Evans Leaders Fund.
\end{acknowledgments}

\vspace{1cm}

The following abbreviations are used in this manuscript:\\

\noindent 
\begin{tabular}{@{}ll}
BZ & Brillouin zone\\
DFT & Density functional theory\\
FD & Finite difference\\
GGA & Generalized gradient approximations\\
HSE & Heyd, Scuseria, and Ernzerhof\\
LDA & Local-density approximation\\
PBE & Perdew, Burke, and Ernzerhof\\
SOC & Spin-orbit coupling\\
VASP & Vienna \textit{ab initio} simulation package\\
XC & Exchange and correlation\\
YSH & Yukawa screened hybrid\\
\end{tabular}

\clearpage

%

%
%

\clearpage

\clearpage

\clearpage

\begin{table}[h]
\caption{Structural and calculation parameters.}\label{tab:WIEN2k paramaters}
\begin{ruledtabular}
\begin{tabular}{lcccc}
\textbf{Parameters}	& \textbf{GaAs} & \textbf{GaN} & \textbf{(CH$_3$NH$_3$)PbI$_3$} & \textbf{1L-MoS$_2$} \\
\hline
Space group & F$\bar{4}$3m (216) & P6$_3$mc (186) & P1 (1) & P$\bar{6}$m2 (187)  \\
Lattice param. ({\AA}) & 5.653~\cite{Averkieva1972} & 3.18, 5.166~\cite{Juza_ZAAC_239_1938} & 6.31~\cite{Dang_CEC_17_2015,Whitfield_SR_6_2016} & 3.16~\cite{Opalovskii_DANS_163_1965}, 29.0\\
& & & (pseudo-cubic) & \\
$R_\text{MT}$ (bohr) & 2.23 (Ga) & 1.90 (Ga) & 0.68 (H) & 2.36 (Mo)\\
& 2.23 (As) & 1.64 (N) & 1.34 (C) & 2.03 (S)\\
& & & 1.26 (N) & \\
& & & 2.50 (Pb) & \\
& & & 2.50 (I) & \\
$n_\text{val}$ & 13 (Ga) & 19 (Ga) & 4 (C) & 14 (Mo)\\
& 15 (As) & 5 (N) & 5 (N) & 6 (S)\\
& & & 18 (Pb) & \\
& & & 17 (I) & \\
$R_{\text{MT}_{\text{min}}}K_{\text{max}}$ & 8.0 & 8.0 & 3.0 & 8.0 \\
$G_{\text{max}}$ & 12 & 12 & 20 & 12 \\
$l_{\text{max}}$ & \multicolumn{4}{c}{10 (all structures)}  \\
$l_{\text{vns}_{\text{max}}}$ & \multicolumn{4}{c}{6 (all structures)}  \\
$k$ mesh & $8\times 8 \times 8$ & $8\times 8 \times 4$ & $3\times 3 \times 3$  & $9\times 9 \times 1$  \\
& ($\Gamma$ centered) & ($\Gamma$ centered) & (shifted)  & ($\Gamma$ centered)  \\
Energy (Ry) and & \multicolumn{4}{c}{$10^{-4}$  (all structures)}  \\
charge converg. & \multicolumn{4}{c}{$10^{-3}$  (all structures)} \\
\end{tabular}
\end{ruledtabular}
\end{table}

\begin{table}[h]
\caption{Length-gauge velocity matrix elements $|v_{mn}^{(x)}|^2$ (at.u.) in GaAs calculated using the finite difference approximations (forward vs. central) Eqs.~\eqref{eq:v_mn length gauge FD forward} and \eqref{eq:v_mn length gauge FD central} with various step sizes $q$. These values are compared with the velocity-gauge momentum matrix elements $|p_{mn}^{(x)}|^2$ (at.u.) from the \texttt{optic} module. The local (GGA-PBE) XC functional is used, which makes the velocity and the length gauges identical. The band degeneracy is given as a superscript in parentheses and the meaning of the subscripts is made clear in Fig .~\ref{fig-GaAs-schematic-band-struct}(a). }\label{tab:v_mn GaAs PBE}
\begin{ruledtabular}
\begin{tabular}{lccc}
\textbf{Transition}	& \multicolumn{2}{c}{$\bm{\sum|v_{mn}^{(x)}|^2}$}	& $\bm{\sum|p_{mn}^{(x)}|^2}$\\
                	& \multicolumn{2}{c}{$\bm{(q= (16.0/3.5/1.2/0.0006)\times 10^{-3}~\textbf{rad~bohr}^{-1})}$}	& \\
                	& \textbf{forward FD}	& \textbf{central FD}	& \\
\hline
$\Gamma_{lh,\,hh}^{(\times 4)} - \Gamma_{c}^{(\times 2)}$		& 0.264/0.402/0.420/0.402			& 0.412/0.422/0.422/0.402			& 0.417\\
$\Gamma_{so}^{(\times 2)} - \Gamma_{c}^{(\times 2)}$		& 0.217/0.202/0.200/0.221			& 0.209/0.201/0.200/0.221	& 0.206\\
\end{tabular}
\end{ruledtabular}
\end{table}

\begin{table}[h] 
\caption{Length-gauge $|v_{mn}^{(x)}|^2$ and velocity-gauge $|p_{mn}^{(x)}|^2$ matrix elements (at.u.) in GaAs calculated using WIEN2k (with YSH) and VASP (with HSE06). Due to the non-local potential the velocity and the length gauges are \textit{not} identical. The band degeneracy is given as a superscript in parentheses and the subscripts are explained in Fig .~\ref{fig-GaAs-schematic-band-struct}(a). The logarithmic deviation between $\sum |p_{vc}^{(x)}|^2$ and $\sum  |v_{vc}^{(x)}|^2$ is given in parentheses ($\Delta$ as per Eq.~\eqref{eq:delta log}).}\label{tab:v_mn GaAs HSE}
\begin{ruledtabular}
\begin{tabular}{lccc}
\textbf{Transition}	& \multicolumn{2}{c}{$\bm{\sum|v_{mn}^{(x)}|^2}$}	& $\bm{\sum|p_{mn}^{(x)}|^2}$\\
                	& \textbf{WIEN2k}	& \textbf{VASP}	& \\
\hline
$\Gamma_{lh,\,hh}^{(\times 4)} - \Gamma_{c}^{(\times 2)}$		& 0.534			& 0.541			& 0.420\\
$\Gamma_{so}^{(\times 2)} - \Gamma_{c}^{(\times 2)}$		& 0.255			& 0.256 & 0.208\\
Total & 0.789 (+23\%) & 0.797 & 0.628 \\
\end{tabular}
\end{ruledtabular}
\end{table}

\begin{table}[h] 
\caption{Length-gauge $|v_{mn}^{(\alpha)}|^2$ and velocity-gauge $|p_{mn}^{(\alpha)}|^2$ matrix elements (at.u.) in GaN calculated using the YSH XC functional.}\label{tab:v_mn GaN HSE}
\begin{ruledtabular}
\begin{tabular}{lcccc}
\textbf{Transition}	& $\bm{\sum|v_{mn}^{(x)}|^2}$ & $\bm{\sum|v_{mn}^{(z)}|^2}$	& $\bm{\sum|p_{mn}^{(x)}|^2}$ & $\bm{\sum|p_{mn}^{(z)}|^2}$\\
\hline
$\Gamma_{hh}^{(\times 2)} - \Gamma_{c}^{(\times 2)}$ 		& 0.211			& 0			& 0.183 & 0\\
$\Gamma_{lh}^{(\times 2)} - \Gamma_{c}^{(\times 2)}$		& 0.256			& 0.055			& 0.163 & 0.042\\
$\Gamma_{so}^{(\times 2)} - \Gamma_{c}^{(\times 2)}$		& 0.026			& 0.507  & 0.018 & 0.377\\
Total & 0.493 (+30\%) & 0.562 (+29\%) & 0.364 & 0.419 \\
\end{tabular}
\end{ruledtabular}
\end{table}

\begin{table}[h]
\caption{Length-gauge $|v_{mn}^{(x)}|^2$ and velocity-gauge $|p_{mn}^{(x)}|^2$ matrix elements (at.u.) in monolayer MoS$_2$ calculated using YSH XC functional.}\label{tab:v_mn MoS2 HSE}
\begin{ruledtabular}
\begin{tabular}{lcc}
\textbf{Transition}	& $\bm{\sum|v_{mn}^{(x)}|^2}$ &  $\bm{\sum|p_{mn}^{(x)}|^2}$ \\
\hline
$K_{v_1} - K_{c_1}$		& 0			& 0 \\
$K_{v_2} - K_{c_1}$		& 0.107			& 0.075 \\
$K_{v_1} - K_{c_2}$		& 0.106			& 0.074 \\
$K_{v_2} - K_{c_2}$		& 0			& 0 \\
Total & 0.213 (+36\%) & 0.149 \\
\end{tabular}
\end{ruledtabular}
\end{table}

\begin{table}[h]
\caption{Length-gauge $|v_{mn}^{(\alpha)}|^2$ and velocity-gauge $|p_{mn}^{(\alpha)}|^2$ matrix elements (at.u.) in pseudo-cubic (CH$_3$NH$_3$)PbI$_3$ calculated using the YSH XC functional.}\label{tab:v_mn MAPbI3 HSE}
\begin{ruledtabular}
\begin{tabular}{lcc}
\textbf{Transition}	& $\bm{\sum|v_{mn}^{(x,y,z)}|^2}$ & $\bm{\sum|p_{mn}^{(x,y,z)}|^2}$\\
\hline
$R_{v}^{(\times 2)} - R_{c}^{(\times 2)}$		& 0.195, 0.150, 0.128 (+11, +11, +8\%)\footnote{The true enhancement should be about 22\% due to inaccuracies in $|p_{mn}^{(x,y,z)}|^2$ values. See text at the end of the Results section for more details.}	& 0.174, 0.135, 0.118 \\
\end{tabular}
\end{ruledtabular}
\end{table}

\end{document}